\documentclass[12pt]{article}
\usepackage{epsfig}
\begin{document}

\begin{flushright}
{\tt hep-th/0305092}
\end{flushright}

\vspace{5mm}

\begin{center}
{{{\Large \bf Tachyon Kinks on Unstable D$p$-branes}}\\[12mm]
{Chanju Kim}\\[1mm]
{\it Department of Physics, Ewha Womans University,
Seoul 120-750, Korea}\\
{\tt cjkim@ewha.ac.kr}\\[6mm]
{Yoonbai Kim,~~O-Kab Kwon,~~Chong Oh Lee}\\[1mm]
{\it BK21 Physics Research Division and Institute of
Basic Science,\\
Sungkyunkwan University, Suwon 440-746, Korea}\\
{\tt yoonbai@skku.ac.kr~~okwon@newton.skku.ac.kr~~cohlee@newton.skku.ac.kr}
}
\end{center}
\vspace{10mm}

\begin{abstract}
In the context of tachyon effective theory coupled to Born-Infeld 
electromagnetic fields, we obtain all possible singularity-free
static flat configurations of codimension one on unstable D$p$-branes.
Computed tension and string charge density suggest that the obtained
kinks are D$(p-1)$ or D$(p-1)$F1-branes.
\end{abstract}


\newpage

\setcounter{equation}{0}
\section{Introduction}

Instability of non-BPS D$p$-brane in string theory is realized by the
presence of of tachyonic mode in its spectrum. It is expected that
condensation of the tachyon takes place and its energetic minimum is the 
closed string vacuum~\cite{oSen}.

Study of tachyon dynamics on a non-BPS D$p$ brane is accomplished in the
scheme of either boundary conformal field theory (BCFT)~\cite{Sen,Sena} or
effective theory with the action~\cite{Sene}--\cite{Klu}
\begin{equation}\label{pfa}
S= -T_{p} \int d^{p+1}x\; V(T) \sqrt{-\det (\eta_{\mu\nu} +
\partial_\mu T\partial_\nu T + F_{\mu\nu})}\, ,
\end{equation}
where $T$ is tachyon and $F_{\mu\nu}$ is the field strength tensor of Abelian
gauge field $A_{\mu}$ on the D$p$-brane.
Since the tachyon potential $V(T)$ measures varying tension, it should satisfy
two boundary values such that $V(T=0)=1$ and $V(T=\infty)=0$.
Specific computation based on (boundary) string field theory~\cite{GS}
gives $V(T)\sim e^{-T^{2}}$ and Ref.~\cite{Senb} suggests $V(T)\sim e^{-T}$ 
for large $T$.\footnote{Comparison of the S-matrix elements from 
string theory and those from effective theory predicts 
$V(T)=e^{-T^{2}}$~\cite{Gar1}.}
Here we adopt a runaway tachyon potential which is convenient 
for analysis~\cite{BLW}--\cite{LLM} and is obtained from open string
theory~\cite{KN,Oku}
\begin{equation}\label{V3}
V(T)=\frac{1}{\cosh \left(\frac{T}{T_{0}}\right)},
\end{equation}
where $T_{0}$ is $\sqrt{2}$ for the non-BPS D-brane in the superstring
and 2 for the bosonic string.

Since the theory of our interest is the effective theory of tachyon
without physical states around its vacuum, an adequate proposal to understand
the tachyon effective action (\ref{pfa}) is made through comparison of
the classical solutions from both the effective theory and the open
string theory which is describable in terms of BCFT~\cite{KN,LLM,SenU}.
The most intriguing solution is so-called rolling tachyon which
provides a real time process of homogeneous tachyon 
configuration and, at late time, becomes a pressureless 
gas~\cite{Sen,Sena,Senb,MZ}. This is understood in terms of
string (field) theory~\cite{ST} and is also a representative
candidate of S-brane~\cite{GSt}--\cite{HHW}. 
In relation with generation of
fundamental string (F1) from unstable D-brane~\cite{Yi,SenU}, 
coupling of Born-Infeld type electromagnetism leads to fluid state of
electric flux tube~\cite{MS,GHY} or that of electromagnetic flux 
tube~\cite{RS,KKKK}.

Spatial inhomogeneity is another important issue~\cite{CKL,Has}, 
particularly in the form of tachyon solitons. In the effective theory of
pure tachyon, tachyon kinks of codimension one have been studied extensively,
however the obtained configurations are either static singular 
solutions~\cite{Klu,MZ2,AIO,LS,Senk} or array of regular 
kink-antikink~~\cite{LLM,KKL,BMS}. Time-dependent kink or periodic 
sinusoidal array is also suffered by encountering of a singularity after
a finite time interval~\cite{CFM}, which appears in a form of blowing-up
energy-momentum tensor for D$p$-branes in bosonic string theory~\cite{Senop}.
Static topological tachyon kink is shown to be BPS D$(p-1)$-brane
with keeping supersymmetry from the study of its worldvolume action of
fluctuations~\cite{Senk,Senc,Send}. Computed tension of the tachyon kink 
coincides with that of D$(p-1)$-brane~\cite{LLM,Senk}. In the context of
string theory, corresponding BCFT description has recently been 
worked out~\cite{SenU}.

Remarkably, introduction of electromagnetic fields regularizes static 
topological kink~\cite{KKL}. For the various constant electric and magnetic
fields, rich spectra of extended tachyon objects of codimension one
are obtained, i.e., they include array of kink-antikink, topological kink,
half-kink, hybrid of two half-kinks, and bounce. 
The corresponding tension and F1 charge density confined on the codimension-one
kink imply that the single unit kink is naturally a candidate of D1 or D1F1. 
When pure electric field is less than or equal to 1, corresponding BCFT 
solutions are also obtained in~\cite{SenU}. 
In the work of Ref.~\cite{KKL}, the obtained kinks
are codimension one objects of unstable D2-brane 
In this paper, we consider unstable D$p$-brane for arbitrary $p$ and find 
all possible static regular tachyon solutions of codimension one, 
which will be interpreted as general flat D$(p-1)$- or D$(p-1)$F1-branes.
Probably static kinks will play an important role in resuming fundamental
dynamical questions, e.g., strings from (rolling) tachyons~\cite{RS2,MV}
and emission of gravitational fields or closed strings~\cite{OS,CLL,LLM,SenU}.

The rest of the paper is organized as follows. In section 2, we analyze
in detail the case of $p=3$. We obtain all possible configurations, 
including array, topological kink, half-kink, hybrid of two half-kinks, 
and bounce. In section 3, we consider the general case of arbitrary
$p$ and show that structure of static regular kinks is the same independent of
$p\ge 2$. We conclude in section 4.

\setcounter{equation}{0}
\section{Tachyon Kinks on Unstable D3-brane}

In this section we analyze the effective tachyon action in (1+3)-dimensions
in detail and find all possible static regular tachyon kink solutions.
We will show that they consist of array of kink-antikink, topological 
kink, half-kink, hybrid of two half-kinks, and bounce
in addition to homogeneous symmetric and broken vacua.
In the context of string theory, the obtained configurations correspond
to D2- and D2F1-branes from unstable D3-brane or to their hybrids.
The solution spectra turn out to be identical to $D_2$ case analyzed
in \cite{KKL}. 

\subsection{Effective theory}

The effective tachyon action for the unstable D3-brane system is
\begin{equation}\label{fa}
S= -T_3 \int d^4x\; V(T) \sqrt{-\det (\eta_{\mu\nu} +
\partial_\mu T\partial_\nu T + F_{\mu\nu})}\, .
\end{equation}
To proceed, we introduce a few notations. We first define
\begin{eqnarray}
X_{\mu\nu}&\equiv & \eta_{\mu\nu} + \partial_\mu T\partial_\nu T + F_{\mu\nu},
\label{Xmn}\\
X&\equiv & \det (X_{\mu\nu}).
\label{X}
\end{eqnarray}
In $X_{\mu\nu}$, we separate barred metric $\bar{\eta}_{\mu\nu}$
and barred field strength tensor $\bar{F}_{\mu\nu}$
\begin{eqnarray}
\bar{\eta}_{\mu\nu}&=&\eta_{\mu\nu}+\partial_{\mu}T\partial_{\nu}T,
\label{emn}\\
\bar{F}_{\mu\nu} &=& F_{\mu\nu}.
\label{fmn}
\end{eqnarray}
Then we have determinant of barred metric
$\bar{\eta}$ and inverse metric
$\bar{\eta}^{\mu\nu}$
\begin{eqnarray}\label{eemn}
\bar{\eta} = -(1 + \partial_\mu T \partial^\mu T),~~~
\bar{\eta}^{\mu\nu} = \eta^{\mu\nu} -
\frac{\partial^\mu T \partial^\nu T}{1
+ \partial_\rho T \partial^\rho T}.
\end{eqnarray}
Contravariant barred field strength tensor $\bar{F}^{\mu\nu}$ and
its dual field strength $\bar{F}^{\ast}_{\mu\nu}$ are
\begin{eqnarray}\label{ffmn}
\bar{F}^{\mu\nu} = \bar{\eta}^{\mu\alpha}
\bar{\eta}^{\nu\beta}F_{\alpha\beta},~~~
\bar{F}^{\ast}_{\mu\nu}= \frac{\bar{\epsilon}_{\mu\nu\alpha\beta}}
{2}
\bar{F}^{\alpha\beta}= \frac{\bar{\epsilon}_{\mu\nu\alpha\beta}}{2}
\bar{\eta}^{\alpha\gamma}\bar{\eta}^{\beta\delta}F_{\gamma\delta},
\end{eqnarray}
where $ \bar{\epsilon}_{\mu\nu\alpha\beta} = \sqrt{-\bar{\eta}}\;
\epsilon_{\mu\nu\alpha\beta}$ with $\epsilon_{0123} = 1$.
In terms of barred quantities
Eq.~(\ref{X}) is computed as
\begin{eqnarray}\label{XX}
X = \bar{\eta} \left[ 1 + \frac12 \bar{F}_{\mu\nu}
\bar{F}^{\mu\nu} - \frac1{16} \left(\bar{F}^*_{\mu\nu}
\bar{F}^{\mu\nu}\right)^2 \right].
\end{eqnarray}

Then equations of motion for the tachyon $T$ and the gauge field
$A_{\mu}$ are
\begin{eqnarray}
\partial_\mu\left( \frac{V}{\sqrt{-X}}
\;C^{\mu\nu}_{\rm S}\; \partial_\nu T\right)
+\sqrt{-X}\; \frac{d V}{d T}
= 0,
\label{te} \\
\partial_\mu\left( \frac{V}{\sqrt{-X}}
\;C^{\mu\nu}_{\rm A}\right) = 0.
\label{ge}
\end{eqnarray}
Here $C^{\mu\nu}_{\rm S}$ and $C^{\mu\nu}_{\rm A}$ are symmetric
and asymmetric part of the cofactor,
\begin{equation}\label{cmn}
C^{\mu\nu} = \bar{\eta}\left(
\bar{\eta}^{\mu\nu} + \bar{F}^{\mu\nu} + \bar{\eta}^{\mu\alpha}
\bar{\eta}^{\beta\gamma}\bar{\eta}^{\delta\nu}
\bar{F}^*_{\alpha\beta}\bar{F}^*_{\gamma\delta}
+\bar{\eta}^{\mu\alpha}\bar{\eta}^{\beta\gamma}
\bar{F}^*_{\alpha\beta}\bar{F}^*_{\gamma\delta}\bar{F}^{\delta\nu}
\right),
\end{equation}
namely,
\begin{eqnarray}
C^{\mu\nu}_{\rm S} &=& \bar{\eta} (
\bar{\eta}^{\mu\nu}
+ \bar{\eta}^{\mu\alpha}
\bar{\eta}^{\beta\gamma}\bar{\eta}^{\delta\nu}
\bar{F}^*_{\alpha\beta}\bar{F}^*_{\gamma\delta}),\\
C^{\mu\nu}_{\rm A} &=& \bar{\eta}(\bar{F}^{\mu\nu}
+\bar{\eta}^{\mu\alpha}\bar{\eta}^{\beta\gamma}
\bar{F}^*_{\alpha\beta}\bar{F}^*_{\gamma\delta}\bar{F}^{\delta\nu}).
\end{eqnarray}
Energy-momentum tensor $T_{\mu\nu}$ in the symmetric form is given by
\begin{eqnarray}\label{tmne}
T^{\mu\nu} =  \frac{T_3 V(T)}{\sqrt{- X}}\;
C_{\rm S}^{\mu\nu},
\end{eqnarray}
where $C_{\mu\nu}\equiv \eta_{\mu\alpha}\eta_{\nu\beta}
C^{\alpha\beta}$.

We denote conjugate momenta of the gauge fields as $\Pi_i$,
\begin{eqnarray}
\Pi_{1}&=&T_{3}\frac{V}{\sqrt{-X}}\left[E_{1}+B_{1}({\bf E}\cdot{\bf B})
\right],
\label{pi1}\\
\Pi_{2}&=&T_{3}\frac{V}{\sqrt{-X}}\left[E_{2}(1+T'^{2})+B_{2}
({\bf E}\cdot{\bf B})\right],\\
\Pi_{3}&=&T_{3}\frac{V}{\sqrt{-X}}\left[E_{3}(1+T'^{2})+B_{3}
({\bf E}\cdot{\bf B})\right].
\label{pi3}
\end{eqnarray}
We only consider the cases of $T = T(x)$, ${\bf E} = {\bf E}(x)$, and
${\bf B} = {\bf B}(x)$ without dependence on $y$ and $z$ coordinates. Then
the equations of motion (\ref{te})--(\ref{ge}) become
\begin{eqnarray}
\partial_1 \left[ \frac{V}{\sqrt{-X}}(1+B_{1}^{2}
-E_{2}^{2}-E_{3}^{2})T^{'} 
\right] &=& \sqrt{-X} \frac{dV}{dT}, 
\label{homeq} \\
\partial_1 \Pi_{1}&=&0,
\label{emzae1} \\
\partial_1 \left\{ \frac{V}{\sqrt{-X}}\left[ -B_{3} +
E_{3}({\bf E}\cdot{\bf B})\right]\right\}  &=& 0,
\label{emzab3} \\
\partial_1 \left\{ \frac{V}{\sqrt{-X}}\left[ -B_{2} +
E_{2}({\bf E}\cdot{\bf B})\right]\right\}  &=& 0,
\label{emzab2}
\end{eqnarray}
where  
\begin{eqnarray}\label{XX1}
-X =\left[1-{\bf E}^{2}+{\bf B}^{2}-({\bf E}\cdot{\bf B})^{2}\right]
+(1+B_{1}^{2}-E_{2}^{2}-E_{3}^{2})T'^{2}.
\end{eqnarray}

{}From Eq.~(\ref{tmne}) we have energy density $\rho$
\begin{equation}\label{chag1}
\rho \equiv T_{00}= T_{3}\frac{V}{\sqrt{-X}}\left[(1+T^{'2})
(1 + B_{1}^{2})+B_{2}^{2}+B_{3}^{2}\right].
\end{equation}
The system in this reference frame carries
nonvanishing linear momentum density 
\begin{equation}\label{mom}
{\cal P}_{i} \equiv T^{0i} = T_{3}\frac{V}{\sqrt{-X}}
\left(\epsilon_{ijk}B^{j}E^{k} - \epsilon_{ij1}B^{1}E^{j}T^{'2}
\right).
\end{equation}
Other nonvanishing components of the energy-momentum tensor
(\ref{tmne}) are
\begin{eqnarray}\label{nvc11}
T_{11} &=& -T_{3}\frac{V(T)}{\sqrt{-X}} 
\left(1+B_{1}^{2}-E_{2}^{2}-E_{3}^{2}\right),\\
\label{nvc22}
T_{22} &=& -T_{3}\frac{V(T)}{\sqrt{-X}} 
\left[(1+T^{'2})(1-E_{3}^{2})-E_{1}^{2}+B_{2}^{2}\right],\\
\label{nvc33}
T_{33} &=& -T_{3}\frac{V(T)}{\sqrt{-X}} 
\left[(1+T^{'2})(1-E_{2}^{2})-E_{1}^{2}+B_{3}^{2}\right],\\
\label{nvc12}
T_{12} &=& -T_{3}\frac{V(T)}{\sqrt{-X}}
\left(E_{1}E_{2}+B_{1}B_{2}\right),\\
\label{nvc13}
T_{13} &=& -T_{3}\frac{V(T)}{\sqrt{-X}}
\left(E_{1}E_{3}+B_{1}B_{3}\right),\\
\label{nvc23}
T_{23} &=& -T_{3}\frac{V(T)}{\sqrt{-X}}
\left[E_{2}E_{3}(1+T^{'2})+B_{2}B_{3}\right].
\end{eqnarray}
Conservation of the energy-momentum tensor, $\partial^{\mu}T_{\mu\nu}=0$,
leads to four constants of motion
\begin{equation}\label{mot}
T_{10}=T_{01},~~T_{11},~~T_{21}=T_{12},~~T_{31}=T_{13}.
\end{equation}

{}From Faraday's law $\partial^{\mu}F^{\ast}_{\mu\nu}=0$,
we find that $E_{2}$, $E_{3}$ and $B_3$ are constants. By an appropriate
choice of coordinates we may assume that, without loss of generality,
\begin{equation}\label{set}
E_{3}=0.
\end{equation}
Then from Eq.~(\ref{emzab3}) we see that
\begin{equation} \label{gam}
\gamma \equiv T_{3}\frac{V(T)}{\sqrt{-X}} 
 = \frac{\Pi_{1}}{E_1 + B_1({\bf E}\cdot {\bf B})}= {\rm constant}.
\end{equation}
Constancy of $T_{11}$ and $T_{13}$, in turn, implies that $B_1$ is also
a constant. Finally, the remaining equations (\ref{emzae1}), (\ref{emzab2})
and (\ref{nvc12}) lead to constancy of $B_2$ and $E_1$.
Therefore ${\bf E}$ and ${\bf B}$ are actually constants.

Substitution of the expression of $X$ (\ref{XX1}) into Eq.~(\ref{gam}) 
summarizes the static system of our interest as a single first-order equation
\begin{equation}\label{beq}
{\cal E}=\frac{1}{2}T'^{2}+U(T).
\end{equation}
Here ${\cal E}$ and $U(T)$ are
\begin{eqnarray}
{\cal E}&=&-\frac{\beta}{2\alpha},
\label{ene}\\
U(T)&=&-\frac{1}{2\alpha\gamma^2}\left[T_{3}V(T)\right]^{2},
\label{upo}
\end{eqnarray}
where $\alpha$ and $\beta$ are defined by
\begin{eqnarray}
\alpha &=& 1 + B_1^2 - E_2^2 ,
\label{alp}\\
\beta &=& 1 - {\bf E}^2 +{\bf B}^{2} - ({\bf E}\cdot {\bf B})^2.
\label{bet}
\end{eqnarray}
Note that every static solution of Eq.~(\ref{beq})
with $x$-dependence alone satisfies
the second-order tachyon equation (\ref{homeq}). 
Though the original system is complicated, we now have a simplified equation
(\ref{beq}) specified by three constants $\alpha$, $\beta$ and $\gamma$.

The choice $E_{3}=0$ allows us to identify additional constants: $T_{02}$, 
$T_{23}$ and $\Pi_3$. Let us summarize the results. The solution space of 
our tachyonic system coupled to Born-Infeld electromagnetism is 
classified by six independent constant parameters; here we choose 
\begin{equation}\label{six}
(\Pi_{1},E_{1},E_{2},B_{1},B_{2},B_{3})
\end{equation}
with $E_{3}=0$.
Then other eight constants $(\Pi_{3},T_{01},T_{02},T_{11},T_{12},T_{13},
T_{23},\gamma)$ are expressed by the above 6 parameters (\ref{six}).
Nontrivial $x$-dependence appears in the quantities
$(\Pi_{2}(x),T_{00}(x),T_{03}(x),T_{22}(x),T_{33}(x))$,
which can be written as
\begin{eqnarray}
&&\Pi_{2} =
\gamma [E_{2}+B_{2}({\bf E}\cdot{\bf B})]+\gamma E_{2} T^{'2},
\label{pi2r}\\
&&T_{00}  =
\gamma(1+{\bf B}^{2})+\gamma(1+B_{1}^{2})T^{'2},
\label{t00p}\\
&&T_{03}  =
\gamma(E_{1}B_{2}-E_{2}B_{1})+\gamma E_{2}B_{1}T^{'2},
\label{t03p}\\
&&T_{22}  =
-\gamma(1-E_{1}^{2}+B_{2}^{2})-\gamma T^{'2},
\label{t22p}\\
&&T_{33}  =
-\gamma(1-E_{1}^{2}-E_{2}^{2}+B_{3}^{2})
-\gamma(1-E_{2}^{2})T^{'2}.
\label{t33r}
\end{eqnarray}
Here the inhomogeneous part $\gamma T'^2$ can also be written as
a constant term plus a piece proportional to square of the tachyon potential,
\begin{equation}
\gamma T'^{2}=-\frac{\beta\gamma}{\alpha}+\frac{1}{\alpha\gamma}
\left[T_{3}V(T)\right]^{2}
\end{equation}
by use of Eq.~(\ref{beq}). 
Note that in the above equations the constant terms are proportional to 
$\Pi_1$ while $(T_2V)^2$ terms are inversely proportional to $\Pi_1$.
Another interesting point is that, for the string charge density $\Pi_2$ 
along $y$-direction, the coefficients of inhomogeneous part is proportional 
to $E_2$. Therefore, the existence of $E_2$ is necessary to achieve a 
confined $F1$ charge on the kink. In addition, the inhomogeneous part of
$T_{33}$ vanishes when $E_2^2=1$.

If we turn off $B_{1}$ and $B_{2}$, the magnetic field has only
$B_{3}$ orthogonal to the electric field ${\bf E}$ so that 
${\bf E}\cdot{\bf B}=0$. Then the system reduces to
the case of unstable D2-brane and subsequently the obtained kink 
configurations are D1- or D1F1-branes \cite{KKL}.

\subsection{Tachyon kink solutions}

\begin{figure}[t]
\centerline{\epsfig{figure=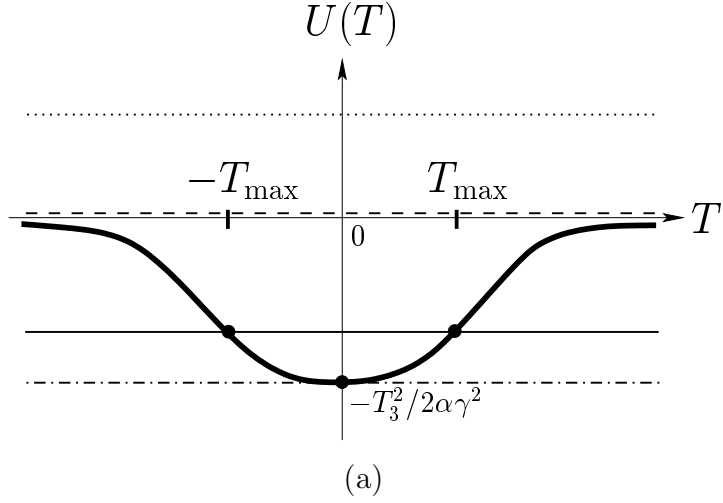,height=60mm}}
\begin{center}
{(a)}
\end{center}
\centerline{\epsfig{figure=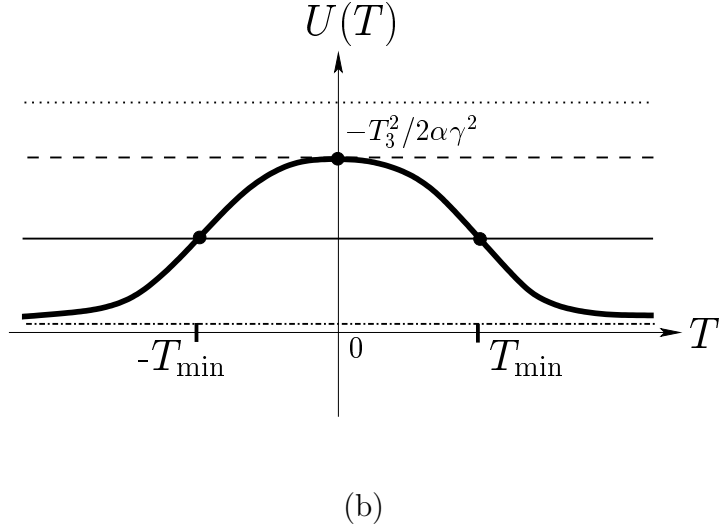,height=60mm}}
\begin{center}
{(b)}
\end{center}
\caption{Two representative shapes of $U(T)$: (a) $\alpha>0$, 
(b) $\alpha<0$.}
\label{fig1}
\end{figure}

In this section we examine the equation (\ref{beq}) and
find all possible regular static kink configurations.
As mentioned previously, each solution is characterized by three parameters
$\alpha$, $\beta$, and $\gamma$ defined in Eqs.~(\ref{alp})--(\ref{gam}).
First of all, from Eq.~(\ref{upo}) we see that the solution space is 
divided into two classes depending on the sign of $\alpha$ since the potential
$U(T)$ flips the sign (see Fig.~\ref{fig1}).
The singular point of $\alpha=0$ will be dealt with separately.

Suppose we have a positive fixed $\alpha$ with a given nonzero $\gamma$.
Then, there are five cases classified by the value of $\beta$ or equivalently
by ${\cal E}$.

(i) When ${\cal E} < U(0)$ ($\beta > T_3^2/\gamma^2$), there exists
no real tachyon solution.

\begin{figure}[t]
\centerline{\epsfig{figure=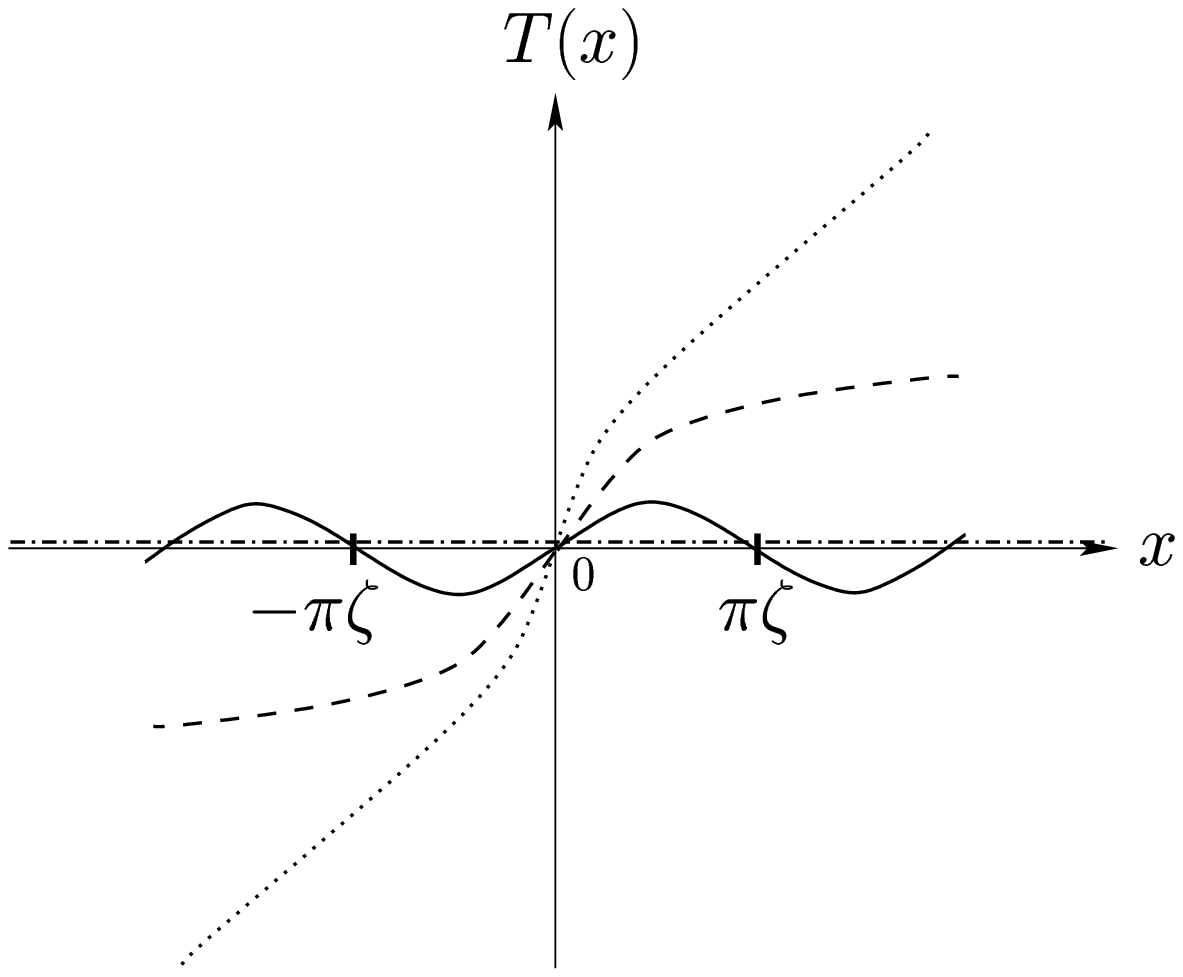,height=80mm}}
\caption{Profiles of tachyon field $T(x)$ for various $\beta$'s
when $\alpha<0$.}
\label{fig2}
\end{figure}

\begin{figure}[t]
\centerline{\epsfig{figure=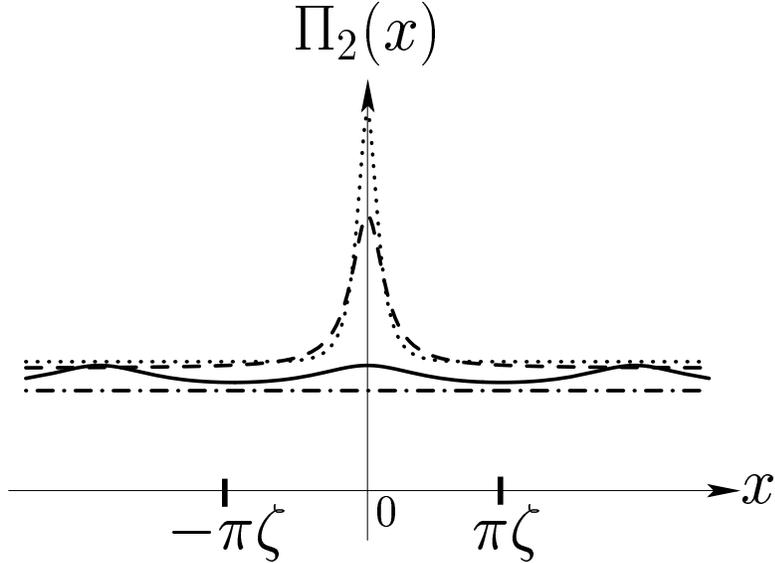,height=80mm}}
\caption{Profiles of string charge density $\Pi_{2}(x)$ when $\alpha>0$.}
\label{fig3}
\end{figure}

(ii) When ${\cal E} = U(0)$ ($\beta = T_3^2/\gamma^2$; 
see the dotted-dashed line
in Fig.~\ref{fig1}-(a)), the constant ontop solution $T(x)=0$ is allowed
(see the dotted-dashed line in Fig.~\ref{fig2}). Correspondingly all 
the quantities
in Eqs.~(\ref{pi2r})--(\ref{t33r}) become constant (see the 
dotted-dashed line in 
Fig.~\ref{fig3}). In the limit of $\Pi_{1}\rightarrow 0$,
$\gamma\rightarrow 0$ and then they vanish.

(iii) When $U(0) < {\cal E} < 0$ ($0<\beta < T_3^2/\gamma^2$),
the tachyon field oscillates between $T_{\rm max}$ and $-T_{\rm max}$ where
$T_{\max}=T_0\,\mbox{arccosh}(T_3/\gamma\sqrt{\beta})$ (see the solid line 
in Fig.~\ref{fig1}-(a)). 
Rewriting Eq.~(\ref{beq}) as an integral equation
\begin{equation}
x = \int^T_0 dT\,\frac{\sqrt{\alpha}}{\sqrt{\beta}
    \sqrt{(T_3 V)^2/\beta\gamma^2-1}},
\end{equation}
we find an exact solution with the tachyon potential (\ref{V3})
\begin{equation}
T(x) = \pm T_0\, \mbox{arcsinh}\left[\sqrt{u^{2}-1} 
          \;\sin \left(\frac{x}{\zeta}\right)\right],
\end{equation}
where period $\zeta$ is
\begin{equation}\label{per}
2\pi \zeta= 2\pi T_0 \sqrt{\frac{\alpha}{\beta}}, 
\end{equation}
and  $u$ is
\begin{equation}\label{u}
u=\frac{T_{3}}{\gamma\sqrt{\beta}}.
\end{equation}

The obtained configuration is a kink (or antikink) which is not
topological. Since the period $\zeta$ (\ref{per}) is finite, space-filling
configuration is an array of kink-antikink 
(see the solid line in Fig.~\ref{fig2}).
The localized part of the quantities in Eqs.~(\ref{pi2r})--(\ref{t33r})
is given by, e.g.,
\begin{eqnarray}\label{pea}
\Pi_{2l}\equiv
\gamma E_2 T'^{2}=
\frac{E_2\beta\gamma}{\alpha}
\frac{1}{-1+\frac{{\rm sec}^{2}(x/\zeta)}{1-(1/u^{2})}}.
\end{eqnarray}
Note that for single kink (antikink) Eq.~(\ref{pea}) 
is peaked at the origin and localized
in the region $-\pi\zeta/2 < x <\pi\zeta/2$ (see the solid line in 
Fig.~\ref{fig3}).
The localized piece of the energy density over the half period
provides the tension of this codimension one object
\begin{eqnarray}
T_{2}&=&\frac{(1+B_{1}^{2})T_{3}^{2}}{\alpha\gamma}
\int^{\frac{\pi}{2}\zeta}_{-\frac{\pi}{2}\zeta}dx \, V^{2}(T(x))\\
&=& \pi T_{0} T_{3}\frac{1+B_{1}^{2}}{\sqrt{\alpha}}.
\label{t21}
\end{eqnarray} 
The multiplicative factor $(1+B_{1}^{2})/\sqrt{\alpha}$ is expected since
the energy density of a Born-Infeld theory increases precisely by this factor
when constant electromagnetic fields are turned on on the worldvolume.
It allows us to interpret this codimension one object as a D2-brane 
in the context of string theory.
Similarly, we have string charge per unit transverse area 
\begin{eqnarray}
Q_{{\rm F}1}&=&\frac{E_{2} T_{3}^{2}}{\alpha\gamma}
\int^{\frac{\pi}{2}\zeta}_{-\frac{\pi}{2}\zeta}dx \, V^{2}(T(x))\\
&=& \pi T_{0} T_{3}\frac{E_{2}}{\sqrt{\alpha}}.
\label{qf1}
\end{eqnarray}
It means that the form of F1's on the D2-brane is a string fluid confined on 
the D2-brane, where string charge is proportional to $E_{2}$. 

In the limit of $\Pi_{1}\rightarrow 0$ or equivalently 
$\gamma \rightarrow 0$
with fixed $\alpha$ and $\beta$, constant piece of 
Eqs.~(\ref{pi2r})--(\ref{t33r}) vanishes and localized part becomes sharply
peaked so that they are given by sums of $\delta$-functions
\begin{eqnarray}
\rho(x) &\simeq& T_{2}
\sum_{n=-\infty}^{\infty}\delta(x-2\pi\zeta),
\label{2r5}\\ 
\Pi_{2}&\simeq& Q_{F1}
\sum_{n=-\infty}^{\infty}\delta(x-2\pi\zeta).
\label{2p5}
\end{eqnarray}
Therefore, the array of kink-antikink is interpreted as that of
infinitely thin D2-$\bar{{\rm D}}$2 or D2F1-$\bar{{\rm D}}$2F1. 
The period $2\pi\zeta$
is unchanged in the singular limit under the tachyon potential of our
consideration (\ref{V3}) but can be changed under a different 
potential~\cite{BMS}.

The obtained array with electromagnetic fields shares the same property 
with the array of pure tachyon kink-antikink except for scaling factor. This 
phenomenon can easily be understood through a rescaling of $x$-coordinate
in the effective action (\ref{fa})
\begin{eqnarray}
S &=& -T_3 \int dt\, dx\, d^{2}x_{\perp}\, V(T)\sqrt{\beta 
        + \alpha\left(\frac{dT}{dx}\right)^2} 
\nonumber \\
&=& - \sqrt{\alpha}\, T_3\int dt\, 
      d\left(T_{0}\frac{x}{\zeta}\right)d^{2}x_{\perp}\, 
      V(T)\sqrt{1 + \left[\frac{d T}{d (T_{0}x/\zeta)}
       \right]^2}.
\label{act3}
\end{eqnarray}
The resultant action (\ref{act3}) is the same as that of pure tachyonic theory
except for the rescaling of $x$-coordinate $x\rightarrow
(\sqrt{\beta/\alpha}\, x)$ and an overall factor
$\sqrt{\alpha}$.

\begin{figure}[t]
\centerline{\epsfig{figure=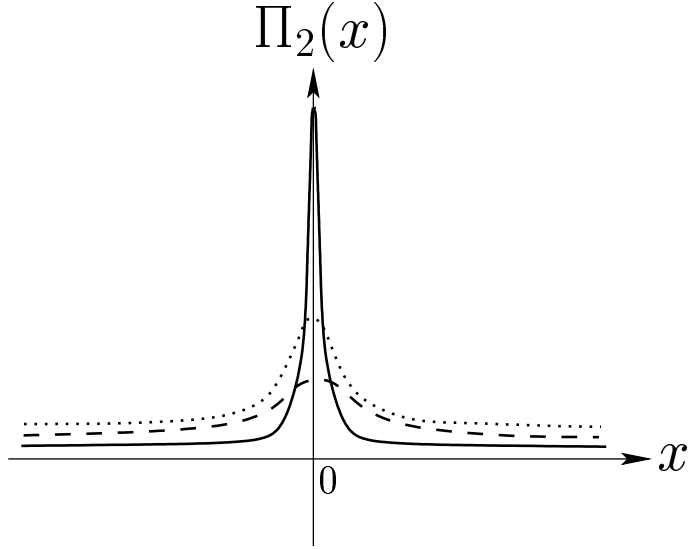,height=80mm}}
\caption{Profiles of string charge density $\Pi_{2}(x)$ for
various $\Pi_{1}$'s. Dashed line with $\Pi_{1}=0.3$, dotted line with 
$\Pi_{1}=0.9$, and solid line with $\Pi_{1}=1.6$.}
\label{fig4}
\end{figure}

(iv) When ${\cal E} = 0$ ($\beta =0$; 
see the dashed line in Fig.~\ref{fig1}-(a)), 
the period of the tachyon kink stretches to infinity, 
$\lim_{\beta\rightarrow 0}2\pi\zeta=
\lim_{\beta\rightarrow 0}2\pi T_0 \sqrt{\alpha/\beta} \rightarrow \infty$.
In addition, $u$ in Eq.~(\ref{u}) diverges with finite ratio
$\zeta/u=\gamma T_{0}\sqrt{\alpha}/T_{3}$.
The solution obtained in this limit is a regular static single 
topological kink configuration with $T'(\pm\infty)=0$
(see the dashed line in Fig.~\ref{fig2})
\begin{equation}
T(x) = T_{0}\, {\rm arcsinh}\left(\frac{ux}{\zeta}
\right).
\end{equation}
The localized piece of various quantities (\ref{pi2r})--(\ref{t33r})
including the energy density and
the string charge density takes Lorentzian shape
(see the dashed line in Fig.~\ref{fig3}) since
\begin{equation}
\gamma T'^{2}=\frac{1}{\alpha\gamma}[T_{3}V(T)]^{2}
=\frac{\pi T_{0}T_{3}}{\sqrt{\alpha}}\frac{\zeta/\pi u}{x^{2}+(\zeta/u)^{2}},
\label{lor}
\end{equation}
where $\zeta/u=(\gamma T_{0}\sqrt{\alpha})/T_{3}$ stands for width of the 
topological kink. When $\Pi_{1}$ goes to zero, the localized piece (\ref{lor})
approaches a $\delta$-function. This sharpening is shown in Fig.~\ref{fig4}.
{}From the coefficients in front of the Lorentzian shape, we read
the tension and the string charge
\begin{equation}\label{t2q}
T_{2}=\pi T_{0} T_{3}\frac{1+B_{1}^{2}}{\sqrt{\alpha}}, \qquad 
Q_{{\rm F}1}=\pi T_{0} T_{3}\frac{E_{2}}{\sqrt{\alpha}}.
\end{equation}

An intriguing point is that the action (\ref{fa}) is rewritten in a 
localized form
\begin{eqnarray}
S&=&-T_{p}\int dt\, dx\,d^{2}x_{\perp}\, 
V(T)\sqrt{\beta+\alpha T'^{2}}
\nonumber\\
&\stackrel{\beta=0}{=}&-(\pm)\sqrt{\alpha}\;
T_{p}\int dt\, dx\, d^{2}x_{\perp}\,V(T) T' \label{exy2}\\
&=&-\int dt\,d^{2}x_{\perp}\,
\sqrt{\alpha}\; T_{p}\int_{-\infty}^{\infty}dT\, V(T),
\label{ext2}
\end{eqnarray}
where $+ (-)$ in the second line (\ref{exy2}) corresponds to the kink 
(the antikink). The exact integral formula for the tachyon field in the
third line (\ref{ext2}) coincides with that of the tension 
\begin{equation}
T_{p-1}=T_{p}\int_{-\infty}^{\infty}dT\, V(T),
\end{equation}
which can be obtained only for the singular limit of the kink in the array 
with or without electromagnetic fields~\cite{Senk}.

(v) When ${\cal E} > 0$ ($\beta < 0$; see the dotted line in 
Fig.~\ref{fig1}-(a)), the solution is given by
(see the dotted line in Fig.~\ref{fig2})
\begin{eqnarray}
T(x) = T_0\, \mbox{arcsinh}\left[\sqrt{1+\bar{u}^{2}} 
          \,\sinh 
\left(\frac{x}{\bar{\zeta}}\right)\right],
\end{eqnarray}
where 
\begin{equation}
\bar{\zeta}=T_{0}\sqrt{-\frac{\alpha}{\beta}}\, ,\qquad \bar{u}^{2}=
-\frac{T_{3}^{2}}{\beta\gamma^{2}}.
\end{equation}
The obtained configuration is also
a topological kink with a finite asymptotic slope 
$T'(\pm \infty)=\pm\sqrt{-(\beta/\alpha)}\ne 0$. Therefore, the localized piece is represented
by $V(T)^{2}$ as
\begin{equation}
\frac{1}{\alpha\gamma}\left[T_{3}V(T)\right]^{2}=
\frac{T_{3}^{2}}{\alpha\gamma}\frac{1}{1+(1+\bar{u}^{2})
\sinh^{2}(x/\bar{\zeta})},
\end{equation}
and then all the quantities in Eqs.~(\ref{pi2r})--(\ref{t33r}) also have
such localized piece in addition to a relatively large constant piece.
As an example, the string charge density $\Pi_{2}$ is plotted 
by the dotted line in Fig.~\ref{fig3}.
Note that the pressure along $z$-direction flips its sign when $E_{2}^{2}>1$.
Similar to the previous case, the tension and the string charge density are
given by
\begin{eqnarray}
T_{2}&=&\frac{(1+B_{1}^{2})T_{3}^{2}}{\alpha\gamma}
\int^{\infty}_{-\infty}dx \, V^{2}(T(x))\nonumber\\
&=&\frac{2 T_{0}T_{3}(1+B_{1}^{2})}{\sqrt{\alpha}}\,{\rm arctan}\left(
\bar{u}\right),
\label{t2e}
\end{eqnarray}
and
\begin{eqnarray}
Q_{{\rm F}1}&=& \frac{E_{2}T_{3}^{2}}{\gamma\alpha}
\int^{\infty}_{-\infty}dx \, V^{2}(T(x))
\nonumber\\
&=& \frac{2 T_{0}T_{3}E_{2}}{\sqrt{\alpha}}\,{\rm arctan}\left(
\bar{u}\right),
\label{q2e}
\end{eqnarray}
which are less than the quantities in Eq.~(\ref{t21}), Eq.~(\ref{qf1}),
and Eq.~(\ref{t2q}). In the limit of divergent $\bar{u}$ with a fixed $\alpha$,
the previous values are reproduced. 
When $\sqrt{-\beta}$ diverges with finite $\bar{u}$ and $\alpha$,
the tachyon kink becomes sharply peaked.

When $\alpha$ is negative ($E_{2}^{2}>1+B_{1}^{2}$), the potential
$U(T)$ is flipped as shown in Fig.~\ref{fig1}-(b) and then character of 
regular static solutions changes drastically. For fixed $\alpha$ and $\gamma$,
the system of our interest is again specified by the value of $\beta$ 
in ${\cal E}$ (\ref{ene}). 
When $\beta$ is positive, the action of our system (\ref{fa}) is rewritten as
\begin{eqnarray}                                                                
S &=& -T_3 \int dt\, dx\, d^2x_{\perp}\, V(T)   
\sqrt{\beta - (-\alpha)\left(\frac{dT}{dx}\right)^2}                            
\nonumber \\                                                                    
&=& - \sqrt{-\alpha}\, T_3\int dt\,  
d\left(\frac{T_0 x}{\bar\zeta}\right)d^2x_{\perp}\,  
V(T)\sqrt{1 - \left[\frac{d T}{d(T_0 x/\bar\zeta)}\right]^2} \, .
\label{rac}
\end{eqnarray}  
But this action (\ref{rac}) is exactly the same as that of rolling tachyon 
which is given by 
\begin{eqnarray} 
S &=& -T_3 \int dt\, dx\, d^2x_{\perp}\, V(T)\sqrt{\beta 
-(1+{\bf B}^{2})\left(\frac{dT}{dt}\right)^2} 
\nonumber \\ 
&=& - \sqrt{(1+{\bf B}^{2})}T_3\int dx\, 
d\left(\frac{T_0 t}{\zeta_{B}}\right)\,d^2x_{\perp}\, 
V(T)\sqrt{1 - \left[\frac{d T}{d (T_0 t/\zeta_{B})} 
\right]^2} 
\end{eqnarray} 
where $\zeta_{B} = \sqrt{(1+{\bf B}^{2})/\beta}$. 
Thus, there exists a one-to-one correspondence between a regular
configuration with spatial $x$-dependence and the time evolution of a 
homogeneous rolling tachyon solution. With this identification, 
the pressure $-T_{11}$ of this system plays the same role as the
Hamiltonian density ${\cal H}$ in the rolling tachyon system.
Since we will find static configurations in a closed form in what follows,
it means that we obtain the most general rolling tachyon solutions in an
arbitrary flat unstable D$p$-brane \cite{Sen,Sena,Senb,MZ,MS,GHY,RS,KKKK}.

\begin{figure}[t]
\centerline{\epsfig{figure=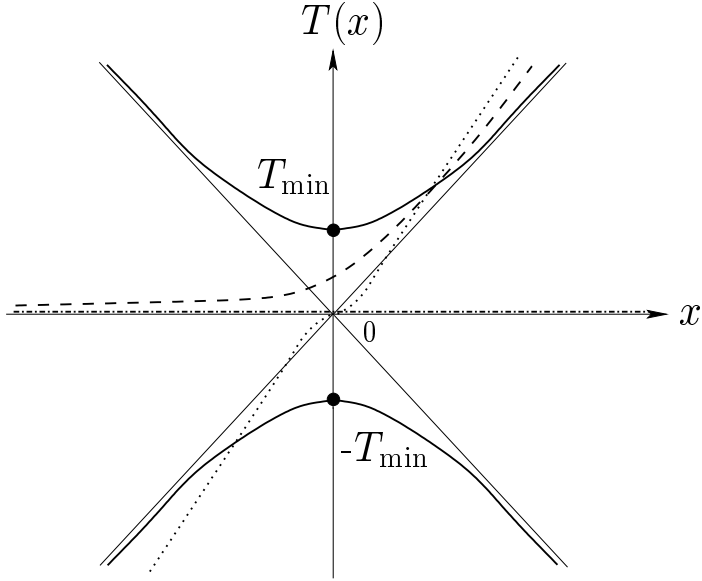,height=80mm}}
\caption{Profiles of tachyon kink and bounce $T(x)$ when $\alpha<0$.}
\label{fig5}
\end{figure}

\begin{figure}[t]
\centerline{\epsfig{figure=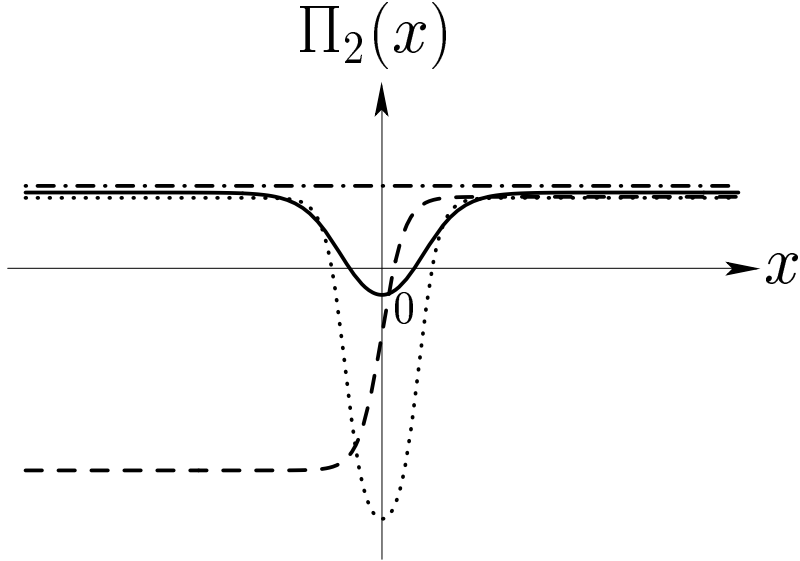,height=80mm}}
\caption{Profiles of string charge density $\Pi_{2}$ when $\alpha>0$.}
\label{fig6}
\end{figure}

(i) When ${\cal E} \to 0^{+}$ ($\beta \to 0^{+}$; see the dotted-dashed
line in Fig.~\ref{fig1}-(b)), constant vacuum solutions, $T(x)=\pm\infty$,
are the only possible configurations 
(see the dotted-dashed line in Figs.~\ref{fig5} and \ref{fig6}). 

(ii) When $0 < {\cal E} < U(0)$ ($0 < \beta < T_{3}^{2}/\gamma^{2}$;
see the solid line in Fig.~\ref{fig1}-(b)), there is a turning point
$T_{\rm min}$ ($-T_{{\rm min}}$) such that 
\begin{equation}
T(x)\ge T_{{\rm min}}=T_{0}\,{\rm arccosh}(u),\qquad
(T(x)\le -T_{{\rm min}}),
\end{equation}
where $u$ is given in Eq.~(\ref{u}).
The corresponding configuration is a bounce
which is convex up (convex down) as shown by the two solid curves in 
Fig.~\ref{fig5}
\begin{equation}\label{tas}
T(x) = T_0\, \mbox{arcsinh}\left[\sqrt{u^{2}-1}
          \,\cosh
\left(\frac{x}{\bar{\zeta}}\right)\right],
\end{equation}
where $\zeta$ is given in Eq.~(\ref{per}).
Its asymptotic slopes are
\begin{equation}
T^{'}(\pm\infty) = \pm T_{0}/\bar{\zeta} ,
\end{equation}
which are shown by the two solid straight lines in Fig.~\ref{fig5}.
Since $\alpha$ is negative, the localized pieces of 
Eqs.~(\ref{pi2r})--(\ref{t33r}) change the sign,
\begin{equation}
\frac{1}{\alpha\gamma}\left[T_{3}V(T)\right]^{2}=
-\frac{T_{3}^{2}}{-\alpha\gamma}\frac{1}{1+(u^{2}-1)
\cosh^{2}(x/\bar{\zeta})},
\end{equation}
which implies negative contribution of the localized energy density $\rho$
and the localized string charge density $\Pi_{2}$ to the constant background 
quantities as shown by the solid line in Fig.~\ref{fig6}.

(iii) When ${\cal E} = U(0)$ ($\beta = T_3^2/\gamma^2$; see the dashed line
in Fig.~\ref{fig1}-(b)), we have the trivial ontop solution $T(x)=0$.
In addition, we find nontrivial tachyon half-kink solutions connecting 
the unstable symmetric vacuum $T(-\infty)=0$ and one of two stable broken
vacua $T(\infty)=\pm\infty$ (see the dashed curve in Fig.~\ref{fig5}),
\begin{equation}
T(x) = \pm T_0\, \mbox{arcsinh}\left[\exp
\left(\frac{x}{\bar{\zeta}}\right)\right].
\end{equation}
Since the half-kink connects smoothly two vacua with different vacuum energy
as $V(T=0)>V(T=\pm\infty)$, the localized piece of 
Eqs.~(\ref{pi2r})--(\ref{t33r}) is monotonically increasing or decreasing
as shown by the dashed line in Fig.~\ref{fig6},
\begin{equation}
\frac{1}{\alpha\gamma}\left[T_{3}V(T)\right]^{2}=
-\frac{T_{3}^{2}}{-\alpha\gamma}
\frac{1}{1+\exp\left(2x/\bar{\zeta}\right)}.
\end{equation}
In the limit of infinite $\Pi_{1}$ with finite $-\alpha$,
$\frac{1}{\alpha\gamma}\left[T_{3}V(T)\right]^{2}$ becomes a step function
with infinite gap.

(iv) When ${\cal E} > U(0)$ ($\beta > T_3^2/\gamma^2$; see the
dotted line in Fig.~\ref{fig1}-(b)), we have
\begin{equation}
T(x) = T_0\, \mbox{arcsinh}\left[\sqrt{1-u^{2}}
          \,\sinh
\left(\frac{x}{\bar{\zeta}}\right)\right].
\end{equation}
Configuration is monotonically increasing (or decreasing) from
$T(-\infty)=\mp\infty$ to $T(\infty)=\pm\infty$ (see the dotted curve in
Fig.~\ref{fig5}) so that this solution can be regarded as hybrid of
two half-kink solutions joined at the origin. Opposite to the similar
kink solutions for positive $\alpha$, slope of the solutions has minimum
value at the origin and maximum value at infinity. Thus the localized piece of
Eqs.~(\ref{pi2r})--(\ref{t33r}) has minimum at the origin due to the flip
of its signature
\begin{equation}
\frac{1}{\alpha\gamma}\left[T_{3}V(T)\right]^{2}=
-\frac{T_{3}^{2}}{-\alpha\gamma}
\frac{1}{1+(1-u^{2})\,\sinh^{2}
\left(x/\bar{\zeta}\right)}.
\end{equation}
The string charge density $\Pi_{2}$ is plotted by the dotted line in
Fig.~\ref{fig6}.

{}Finally we consider the case $\alpha=0$. If we multiply
$\alpha$ to Eq.~(\ref{beq}) and take the limit of $\alpha\rightarrow 0$,
then $T'^{2}$ term disappears. The original tachyon equation
(\ref{homeq}) reduces to $dV/dT=0$ so that we only have homogeneous
vacuum solutions, $T(x)=0$ or $T(x)=\pm\infty$.

\setcounter{equation}{0}
\section{Tachyon Kinks on Unstable D$p$-brane}
The analysis in the previous section can be applied to general
unstable
D$p$-branes without much difficulty. In this section we will
show that, even for general D$p$-branes, the equations of motion reduce
to a single first-order equation of the form in Eq.~(\ref{beq}) for static
case with inhomogeneity in the $x$-direction. Thus what we have
obtained in the previous section and also in Ref.~\cite{KKL} are actually
the most general type of regular static kink solutions of codimension one
for D$p$-branes.

Assuming only the $x(=x^1)$-dependences in the fields $T$ and $A_{\mu}$,
the determinant $X$ in the action (\ref {pfa}) can be written as
\begin{eqnarray} \label{dpx}
-X &=& -\det (\eta_{\mu\nu} +\partial_\mu T\partial_\nu T + F_{\mu\nu})
\nonumber \\
   &=&  \beta_{p}+ {T'}^2 \alpha_{p},
\end{eqnarray}
where
\begin{equation}
\beta_{p} = -\det (\eta_{\mu\nu} + F_{\mu\nu}), 
\end{equation}
and $\alpha_{p}=C^{11}$ is the 11-component of the cofactor of $X_{\mu\nu}$.
The equations of motion are
\begin{eqnarray} \label{dpeom}
& & \partial_1\left(T_p\frac{V}{\sqrt{-X}}C^{11} T'\right) 
     = - \sqrt{- X} T_p\frac{dV}{dT}, \nonumber\\
& & \partial_1 \Pi_1 = 0, \nonumber\\
& & \partial_1\left(T_p\frac{V}{\sqrt{-X}} C^{1i}_{\rm A}\right) = 0,
\qquad (i=2,3,\ldots,p-1),
\end{eqnarray}
where
\begin{equation} \label{dppi1}
\Pi_1 = T_p \frac{V}{\sqrt{-X}}C^{01}_{\rm A}.
\end{equation}
The expression of the energy-momentum tensor is
again given by the symmetric part of the cofactor
as in the D3-brane, namely,
\begin{equation}
T^{\mu\nu} = T_p\frac{V}{\sqrt{-X}}~ C^{\mu\nu}_{\rm S}.
\end{equation}
Then the conservation equation, $\partial_\mu T^{\mu\nu}=0$, becomes
\begin{equation} \label{dpemc}
\partial_1 T^{1\mu} 
  = \partial_1\left(T_p\frac{V}{\sqrt{-X}} C^{1\mu}_{\rm S} \right) =0.
\end{equation}
Addition of Eq.~(\ref{dpeom}) and Eq.~(\ref{dpemc}) leads to
\begin{equation}\label{dpeq}
  \partial_1\left(T_p\frac{V}{\sqrt{-X}} C^{\mu1} \right) =0.
\end{equation}

As in the case of unstable D3-brane,
the Bianchi identity $\partial_{(\mu} F_{\nu\lambda)}=0$
imposes $(p-1)(p-2)/2$ constraints on the $p(p-1)/2$ components of
field strength tensor with our ansatz:
$E_k=\mbox{constant}$ and $F_{kl}=\mbox{constant}$, where $k \neq1$ and
$l \neq 1$. Now we argue that, in fact, all the field strength components are
constant including $E_1$ and $F_{1\mu}$.
Since the cofactor $C^{11}$ in Eq.~(\ref{dpx})
does not contain $F_{\mu\nu}$
with $\mu=1$ or $\nu=1$, Eq.~(\ref{dpeq}) implies that
\begin{equation}\label{dpgamma}
  \gamma_{p}\equiv T_p\frac{V}{\sqrt{-X}} =\mbox{constant},
\end{equation}
as in Eq.~(\ref{gam}).
Now the only remaining nontrivial equations are, from Eq.~(\ref{dpeq}),
\begin{eqnarray}
&&\partial_1 C^{01} = 0, \nonumber \\
&&\partial_1 C^{k1} = 0, \qquad (k = 2,3,\ldots,p-1).
\end{eqnarray}
However, note that these equations are actually homogeneous coupled
linear equations of $\partial_1 F_{1\mu}$ since each $X_{1\nu}$ ($\nu\neq 1$)
appears precisely once in every term of $C^{\mu1}$. Therefore, as long as
the determinant made of the coefficients of $\partial_1 F_{1\mu}$ does
not vanish, we have $F_{\mu\nu} = \mbox{constant}$ for all
$\mu,\ \nu$. (When the determinant does vanish, we can treat the case
in a similar fashion as done in section 2.)

Combining Eqs.~(\ref{dpx}) and (\ref{dpgamma}), we again obtain the
single first-order equation (\ref{beq}) with
\begin{eqnarray} \label{dpupo}
{\cal E}&=&-\frac{\beta_{p}}{2\alpha_{p}}, \nonumber \\
U(T)&=&-\frac{1}{2\alpha_{p}\gamma_{p}^{2}}[T_{p}V(T)]^{2}.
\end{eqnarray}
Then, the rest of the analysis is the same as in D3 case, e.g.,
the solution space for regular static kinks of codimension one
is classified by three constants $\alpha_{p}$, $\beta_{p}$,
and $\gamma_{p}$ and so on.

When $p=1$, the field strength tensor has only one component of electric 
field and the solutions involve only those of $\alpha_p >0$ case.

\setcounter{equation}{0}
\section{Conclusion}
In this paper we have analyzed regular static solutions of
codimension-one extended objects in the effective theory of a real tachyon,
described by Born-Infeld type action with a runaway tachyon potential
coupled to an Abelian gauge field. On arbitrary flat unstable D$p$-brane,
the types of codimension-one extended objects are the same
irrespective of $p$, $(p\ge 2)$. The static regular kink-type solutions
on unstable D$p$-brane are shown to be
classified by three parameters:
$\beta_{p} = -\det(\eta_{\mu\nu}+F_{\mu\nu})$,
$\alpha_p = C^{11}$ (11-component of the cofactor of
$V_{\mu\nu}=\eta_{\mu\nu}+\partial_{\mu}T\partial_{\nu}T+F_{\mu\nu}$) and
$\gamma_p = T_p V/\sqrt{-X}$.
Detailed analysis has been carried out for D3 case.
Species of the obtained solutions are summarized
in Table~1 for various $\alpha_{p}$ and $\beta_{p}$ with
fixed nonzero $\gamma_{p}$.
\vspace{2mm}
\begin{center}
\renewcommand{\arraystretch}{1.4}
\begin{tabular}{|c | c | c|} \hline
 & $\alpha_{p}>0$ & $\alpha_{p}<0$ \\ \hline
$\beta_{p} <0$ & topological kink with $T'(\pm\infty)\ne 0$ &  \\
$\beta_{p}=0$  & topological kink & constant vacuum, $T=\pm\infty$ \\
$0<\beta_{p} <1/\gamma_{p}$ & array of kink-antikink & bounce \\
$\beta_{p}=T_{3}^{2}/\gamma_{p}^{2}$ & constant ontop, $T=0$ & constant ontop,
$T=0$, $\&$ half-kink \\
$\beta_{p}>T_{3}^{2}/\gamma_{p}^{2}$  & & hybrid of two half-kinks \\ \hline
\end{tabular}
\end{center}
\begin{center}{
Table 1: List of regular static configurations.}
\end{center}
\vspace{2mm}

For the single unit object listed in the left column of Table 1
($\alpha_{p}>0$), the tension of lower dimensional branes is
correctly reproduced. When the electric field along the kink direction
is nonzero, the fundamental string charge per unit $(p-1)$-dimensional
transverse volume has a confined piece. This suggests that it may be
interpreted as D$(p-1)$- or D$(p-1)$F1-brane on the unstable D$p$-brane.
If $\alpha_p<0$, due to the correspondence between the static case and 
time-dependent rolling tachyon case, the solutions found here may also 
be interpreted as the most general homogeneous rolling tachyon solutions
of an arbitrary flat D$p$-brane.

\section*{Acknowledgements}
This work was supported by Korea Research Foundation Grant
KRF-2002-070-C00025(C.K.) and is
the result of research activities (Astrophysical Research
Center for the Structure and Evolution of the Cosmos (ARCSEC))
supported by Korea Science $\&$ Engineering Foundation(Y.K. and O.K.).

\end{document}